\documentclass[twocolumn,hyperpdf,amsmath,amssymb,aps,prd,10pt,superscriptaddress,nofootinbib,noeprint,preprintnumbers,floatfix]{revtex4-1}

\usepackage{graphicx, color}
\graphicspath{{./figures/}}

\usepackage[letterspace=-10]{microtype} 

\usepackage{bm, amsmath, amsfonts, amssymb}
\usepackage{multirow, tabularx, dcolumn}
\usepackage{mathtools,leftidx}
\usepackage{blkarray}

\usepackage[utf8]{inputenc} 
\usepackage{hyperref}

\definecolor{jlab_red}{RGB}{192,39,45}
\definecolor{jlab_orange}{RGB}{249,102,0}
\definecolor{jlab_blue}{RGB}{47,122,121}
\definecolor{jlab_green}{RGB}{65,125,10}
\definecolor{jlab_gray}{gray}{0.6}

\newcommand{\cm}{\ensuremath{\mathsf{cm}}}

\newcommand{\SlJ}{\prescript{2S+1\!}{}{\ell}_J}

\newcommand{\threeSone}{\prescript{3\!}{}{S}_1} 
\newcommand{\threePzero}{\prescript{3\!}{}{P}_0}
\newcommand{\threePone}{\prescript{3\!}{}{P}_1}
\newcommand{\threePtwo}{\prescript{3\!}{}{P}_2}  
\newcommand{\threeDone}{\prescript{3\!}{}{D}_1}

\newcommand{\onePone}{\prescript{1\!}{}{P}_1} 
\newcommand{\oneFthree}{\prescript{1\!}{}{F}_3} 

\newcommand{\fiveStwo}{\prescript{5\!}{}{S}_2}  
\newcommand{\oneDtwo}{\prescript{1\!}{}{D}_2}  
\newcommand{\fiveDtwo}{\prescript{5\!}{}{D}_2}  
\newcommand{\fiveDfour}{\prescript{5\!}{}{D}_4}  
\hypersetup{%
pdftitle = {eigenvalue root finding},
pdfsubject = {QCD},
pdfkeywords = {QCD, Hadron, Physics, Lattice, Meson, Scattering},
pdfauthor = {Hadron Spectrum Collaboration},
colorlinks = {true},
filecolor = {black},
linkcolor = {jlab_blue},
menucolor = {black},
citecolor = {jlab_green},
urlcolor = {jlab_green},
}{}

\begin{document}

\preprint{JLAB-THY-20-3133}

\title{ Efficient solution of the multi-channel L\"uscher determinant condition through eigenvalue decomposition}

%%%%%%%%%%%%%%%%%%%%%%%%%%%%%%%%%%%%%%%%%%%%%%%%%%%%%%%%%%%%%%%%%%%%%%
\author{Antoni~J.~Woss}
\email{a.j.woss@damtp.cam.ac.uk}
\affiliation{DAMTP, University of Cambridge, Centre for Mathematical Sciences, Wilberforce Road, Cambridge, CB3 0WA, UK}
%%%%%%%%%%%%%%%%%%%%%%%%%%%%%%%%%%%%%%%%%%%%%%%%%%%%%%%%%%%%%%%%%%%%%%
\author{David~J.~Wilson}
\email{d.j.wilson@damtp.cam.ac.uk}
\affiliation{DAMTP, University of Cambridge, Centre for Mathematical Sciences, Wilberforce Road, Cambridge, CB3 0WA, UK}
%%%%%%%%%%%%%%%%%%%%%%%%%%%%%%%%%%%%%%%%%%%%%%%%%%%%%%%%%%%%%%%%%%%%%%
\author{Jozef~J.~Dudek}
\email{dudek@jlab.org}
\affiliation{\lsstyle Thomas Jefferson National Accelerator Facility, 12000 Jefferson Avenue, Newport News, VA 23606, USA}
\affiliation{Department of Physics, College of William and Mary, Williamsburg, VA 23187, USA}
%%%%%%%%%%%%%%%%%%%%%%%%%%%%%%%%%%%%%%%%%%%%%%%%%%%%%%%%%%%%%%%%%%%%%%

\collaboration{for the Hadron Spectrum Collaboration}
\date{\today}

%%%%%%%%%%%%%%%%%%%%%%%%%%%%%%%%%%%%%%%%%%%%%%%%%%%%%%%%%%%%%%%%%%%%%%
\begin{abstract}
We present a method for efficiently finding solutions of L\"uscher's quantisation condition, the equation which relates two-particle scattering amplitudes to the discrete spectrum of states in a periodic spatial volume of finite extent such as that present in lattice QCD. The approach proposed is based on an eigenvalue decomposition in the space of coupled-channels and partial-waves, which proves to have several desirable and simplifying features that are of great benefit when considering problems beyond simple elastic scattering of spinless particles. We illustrate the method with a toy model of vector-vector scattering featuring a high density of solutions, and with an application to explicit lattice QCD energy level data describing $J^P=1^-$ and $1^+$ scattering in several coupled channels. 
\end{abstract}
%%%%%%%%%%%%%%%%%%%%%%%%%%%%%%%%%%%%%%%%%%%%%%%%%%%%%%%%%%%%%%%%%%%%%%

\maketitle

\section{Introduction}\label{SecI}

Hadron spectroscopy is enjoying a renaissance driven by a global experimental program producing new and unexpected excited states. Excited states are observed as resonant enhancements in hadron-hadron \emph{partial-wave scattering amplitudes}, a common quantity appearing in both experiment and theory that allows these resonances to be understood from the fundamental equations governing hadron interactions, Quantum Chromodynamics (QCD).

One approach for making predictions from QCD is \emph{lattice QCD}, in which a finite Euclidean spacetime is discretized and the path integrals of the quantum field theory are sampled numerically using Monte-Carlo techniques. This approach makes only controlled and improvable approximations, and enables correlation functions and thus the finite-volume spectrum to be computed. Provided sufficiently large volumes and sufficiently small lattice spacings are used, reliable predictions can be made. 

The past decade has seen the state of the art in lattice QCD calculations move to consider scattering systems of ever increasing complexity, from elastic cases like $\pi\pi$~\cite{Sharpe:1992pp,Beane:2007xs,Sasaki:2008sv,Feng:2009ij,Dudek:2010ew,Beane:2011sc,Dudek:2012gj,Dudek:2012xn,Bai:2015nea,Blum:2015ywa,Helmes:2015gla,Liu:2016cba,Briceno:2016mjc,Bulava:2016mks,Alexandrou:2017mpi,Andersen:2018mau}, to the first calculations of coupled-channel scattering~\cite{Dudek:2014qha, Wilson:2014cna, Wilson:2015dqa, Dudek:2016cru, Cheung:2016bym}, to recent studies of multiple coupled channels such as $\pi \pi, K\bar{K}, \eta \eta$ considering both $J^P=0^+$ and $2^+$ which resonate at low energies~\cite{Briceno:2017qmb}, and systems which feature scattering hadrons of non-zero spin, such as $\pi \omega$~\cite{Lang:2014tia, Woss:2019hse} or $\pi N$~\cite{Torok:2009dg, Lang:2012db, Detmold:2015qwf, Lang:2016hnn, Andersen:2017una}.

In these calculations, hadron-hadron scattering amplitudes are inferred from their effect on the discrete spectrum of states in a periodic spatial volume, usually cubic, defined by the lattice. The precise relationship is encoded in the L\"uscher quantization condition \cite{Luscher:1990ux,Rummukainen:1995vs,Bedaque:2004kc,He:2005ey,Kim:2005gf,Fu:2011xz,Leskovec:2012gb,Gockeler:2012yj,Bernard:2010fp,Doring:2012eu,Hansen:2012tf,Briceno:2012yi,Guo:2012hv,Briceno:2014oea}, which can be written as\,\footnote{This is valid for any two-hadron to two-hadron scattering process -- significant recent steps towards a general quantization condition relevant for three-hadron channels have been made~\cite{Hansen:2014eka,Briceno:2017tce,Hammer:2017kms,Hammer:2017uqm,Mai:2017bge,Doring:2018xxx,Briceno:2018aml,Hansen:2019nir,Briceno:2019muc,Romero-Lopez:2019qrt}.}
\begin{align}\label{Intro:Eq:1+irt}
\!\!\det \big[ &\bm{D}(E_\cm) \big]  =0 \,, \;\;\;\text{where} \nonumber \\[0.8ex]
&\bm{D}(E_\cm)  =\bm{1}+i\bm{\rho}(E_\cm) \!\cdot\! \bm{t}(E_\cm) \!\cdot\!  \big(\bm{1} \!+\!  i\bm{\mathcal{\bm{M}}}(E_\cm,L)   \big) .
\end{align}
In this expression, $\bm{\mathcal{\bm{M}}}$ is a matrix of known kinematic factors that encodes the dependence on the $L \!\times\! L \!\times\! L$ volume, \mbox{$\bm{\rho}$ is a} diagonal matrix of phase-space factors, and $\bm{t}$ is the $t$-matrix describing scattering, related to the $S$-matrix by $\bm{S}=\bm{1}+2i\, \sqrt{\bm{\rho}}\cdot \bm{t} \cdot \!\!\sqrt{\bm{\rho} }$. The matrix indexing runs over hadron-hadron channels and partial-waves, with a version of Eqn.~\ref{Intro:Eq:1+irt} existing for each irreducible representation ({\em irrep}) of the relevant symmetry group. Because the symmetry of the cubic boundary is reduced with respect to the full rotation group, each irrep features multiple partial-waves\,\footnote{Formally an infinite number of partial-waves are present, but higher angular momenta are suppressed at low energies by barrier factors and may be neglected in practice. Explicit relations for the \emph{subduction} in the case of systems with net momentum can be found in \cite{Thomas:2011rh}.}. The space of hadron-hadron scattering channels is determined by those channels which are kinematically accessible in the energy region of interest.

The complex scalar function $\det \big[ \bm{D}(E_\cm) \big]$ features not only \emph{zeroes} corresponding to the finite-volume spectrum, but also \emph{divergences} which originate in the matrix $\bm{\mathcal{M}}$, that appear at each energy where a hadron-hadron pair could go on-shell in a theory without interactions, i.e. where the individual hadrons have allowed finite-volume momenta, $\tfrac{2\pi}{L}(n_x, n_y, n_z)$ with $n_i \in \mathbb{Z}$.

Considering Eqn.~\ref{Intro:Eq:1+irt}, for a given scattering matrix, $\bm{t}(E_\cm)$, one can determine the finite-volume spectrum by finding all zeroes in a desired energy region, and in practice this is done numerically using root-finding algorithms applied to the scalar function $\det \big[ \bm{D}(E_\cm) \big]$. This requires computation of the elements of the matrix $\bm{\mathcal{M}}$ which comes with some non-negligible computational overhead, so approaches are generally preferred that evaluate the function less often. It is common to encounter cases where hadron-hadron interactions are relatively weak in some energy regions, such that zeroes lie very close to non-interacting energies where $\det \big[ \bm{D}(E_\cm) \big]$ diverges. This can pose a challenge to conventional approaches to root-finding which rely upon bracketing a zero by locating function values of opposite sign either side of the zero. 

The {\em hadspec} collaboration has pioneered an approach for determining \emph{coupled-channel} scattering amplitudes in which the energy dependence of the $t$-matrix is parameterized, and the best description of the lattice QCD computed finite-volume spectrum is found by varying the free parameters in the scattering amplitude. In practice this requires finding the solutions of Eqn.~\ref{Intro:Eq:1+irt} for each considered volume and irrep, at every iteration of the scattering amplitude parameters which are being varied under a $\chi^2$ minimization, where the $\chi^2$ measures the difference between the found `model' spectrum (the zeroes) and the lattice spectrum. In this approach it is vitally important to find {\em all} the relevant solutions at every iteration, as if any are missed in any iteration, the sampling of the $\chi^2$ surface will feature discontinuous jumps, which are likely to cause failure in the minimization routine.

A more fundamental challenge arises when stable particles with nonzero spin feature in the scattering process. In these cases it becomes possible for there to be zeroes of $\det \big[ \bm{D}(E_\cm) \big]$ with multiplicity larger than one~\cite{Woss:2018irj,Woss:2019hse}, which can appear as the function {\em touching} zero rather than {\em crossing} zero, as here bracketing will fail altogether. If the zero is found by some non-bracketing approach~\cite{Morningstar:2017spu}, it is generally not immediately clear what its multiplicity is. As interactions are typically of different strengths in different partial-waves, this degeneracy of zeroes is broken, but this leads to multiple zeroes appearing in a very small energy range. This remains a challenge for bracketing approaches: for example, suppose two zeroes are separated by a small energy, and the bracketing `grid' samples points such that the two closest neighbouring points lie below the first zero and above the second zero, no change of sign will be observed, and no hint provided that any zeroes are present. 

In this paper, we will present a pragmatic approach to finding zeroes of the determinant condition that is both computationally fast and reliable. The approach makes use of the eigenvalue decomposition of a suitably transformed version of the matrix, $\bm{D}$, appearing under the determinant in Eqn.~\ref{Intro:Eq:1+irt}, using the well-known result that a zero of $\det \bm{D}$ corresponds to at least one eigenvalue of $\bm{D}$ taking a zero value. In practice we find that considering the energy-dependence of the eigenvalues reduces an \mbox{$n$-channel} problem to one which is no more difficult to solve than $n$ independent one-channel problems, where bracketing approaches typically work well. 

Several considerations are made in search of a practical implementation. Firstly, whether there are transformations on $\bm{D}$ defined in Eqn.~\ref{Intro:Eq:1+irt} that preserve the zeroes of $\det \bm{D}$ while also simplifying the numerical problem of finding the zeroes. Secondly, how one should match the eigenvalues between neighbouring energy steps to ensure the $\lambda_p(E_\cm)$ are continuous functions, and finally how to handle isolated special points like kinematic thresholds, and energies at which the eigenvalues diverge.

In the remainder of the paper, after describing our implementation, we will illustrate the method using a toy model of scattering of two vector mesons featuring very dense zeroes, before presenting results from an explicit lattice QCD calculation of $I\!=\!1$, $G\!=\!+$ scattering of $\pi\pi$, $K\overline{K}$, $\pi \omega$ and $\pi\phi$, where both $J^P=1^+$ and $J^P=1^-$ amplitudes appear.

\section{Eigenvalue Decomposition}\label{SecII}

We consider Eqn.~\ref{Intro:Eq:1+irt} where we have truncated the space of partial-waves and hadron-hadron scattering channels to some finite number, rendering $\bm{D}(E_\cm)$ an $\mathfrak{n} \times \mathfrak{n}$ matrix whose determinant can be decomposed as the product of its $\mathfrak{n}$ eigenvalues $\lambda_p(E_\cm)$, 
\begin{align*}
\det\left[\bm{D}(E_\cm)\right] = \prod_{p=1}^\mathfrak{n} \lambda_p(E_\cm),
\end{align*}
where
\begin{equation*}
\bm{D}(E_\cm)  \, \bm{v}^{(p)}(E_\cm) = \lambda_p(E_\cm) \,  \bm{v}^{(p)}(E_\cm)\, ,
\end{equation*}
defines the corresponding eigenvectors\,\footnote{A related approach is proposed in the context of the three-body quantization condition in Ref.~\cite{Blanton:2019igq}.}. Solutions of $\det\left[{\bm{D}}(E_\cm)\right]=0$ occur when at least one eigenvalue takes a zero value, so our approach will be to find each eigenvalue as a function of energy and then perform \mbox{root-finding} on each independently. In order to do this, the eigenvalues must be reliably matched at each step in energy over the entire range considered, otherwise `jumps' due to mismatching could be mistaken for true zeroes and vice-versa, true zeroes could be missed.

For the form of the quantization condition given in Eqn.~\ref{Intro:Eq:1+irt}, at least one of the eigenvalues diverges at every non-interacting energy owing to a pole in $\bm{\mathcal{M}}$, originating in the L\"uscher zeta-functions\,\footnote{See Eq.~\ref{AppA:Eq:c} in Appendix~\ref{AppA}.}. Such a divergence can present a problem when attempting to match the eigenvalues or eigenvectors at energies close to these divergences -- the assumption that a small change in energy leads to a small perturbation of $\bm{D}(E_\cm)$ is invalid near the non-interacting energies as the matrix is unbounded. In those scattering processes where interactions are relatively weak, it is very common that solutions of the quantization condition lie in a neighbourhood of these non-interacting energies, so removing these divergences is a priority.

Fortunately it is quite straightforward to transform $\bm{D}(E_\cm)$ in such a way as to remove the non-interacting divergences while leaving the zeroes of $\det \big[ \bm{D}(E_\cm) \big]$ unchanged. For example,
\begin{align}\label{SecII:Eq:1+irt_1+sv}
\bm{D} &= \bm{1}+i\bm{\rho}\cdot \bm{t}\cdot  (  \bm{1}+i\bm{\mathcal{\bm{M}}} ) \nonumber \\
&= \tfrac{1}{2}\, \bm{\rho}^{\frac{1}{2}} \cdot 
\big(\bm{1} \!+\! \bm{S} \!\cdot\! \bm{V}\big) \cdot
\big(\bm{1} \!-\! i\bm{\mathcal{M}}\big) \cdot
 \bm{\rho}^{-\frac{1}{2}}
\, ,
\end{align} 
where $\bm{V}(E_\cm, L)=(\bm{1}+i\bm{\mathcal{M}})(\bm{1}-i\bm{\mathcal{M}})^{-1}$. The divergences at non-interacting energies have been explicitly factorised off in $(\bm{1}-i\bm{\mathcal{M}})$, leaving $\bm{V}$ which is free of these divergences. It follows that the zeroes of $\det \big[ \bm{D}(E_\cm) \big]$ are also zeroes of $\det\big[\bm{{D_V}}(E_\cm)\big]$, where
\begin{align}\label{SecII:Eq:1+sv}
\bm{D_V}(E_\cm) = \bm{1}+\bm{S}(E_\cm)\cdot \bm{V}(E_\cm,L)\,.
\end{align}
A closely related alternative\,\footnote{Which appeared originally in the elastic case in Ref.~\cite{Luscher:1990ux}.} is
\begin{align}\label{SecII:Eq:s+u}
\det\left[\bm{D_U}(E_\cm)\right]=\det\left[\bm{S}(E_\cm) + \bm{U}(E_\cm,L)\right]=0 \,,
\end{align}
where $\bm{U}=(\bm{1}-i\bm{\mathcal{M}})(\bm{1}+i\bm{\mathcal{M}})^{-1} = \bm{V}^{-1}$, which is also finite at non-interacting energies.
Note that like $\bm{\mathcal{M}}$, both $\bm{V}$ and $\bm{U}$ are diagonal in hadron-hadron channel but dense in partial-wave space.

The properties of the matrices $\bm{S}$, $\bm{U}$ and $\bm{V}$ depend upon energy, and in particular whether any channel is kinematically closed at any given energy. We begin by considering the case of energies high enough that all included channels are kinematically open.

%%%%%%%%%%%%%%%%%%%%%%%%%%%%%%%%%%%%%%%%%%%%%%%%%%%%%%%%%%%%%%%%%%%%%%%%%%%%%%%%%%%%%%%%%%%%%%%%%%%%%%%%%%%%
\subsection{With all channels open}\label{SecII:A}
%%%%%%%%%%%%%%%%%%%%%%%%%%%%%%%%%%%%%%%%%%%%%%%%%%%%%%%%%%%%%%%%%%%%%%%%%%%%%%%%%%%%%%%%%%%%%%%%%%%%%%%%%%%%

%
Consider a scattering system with $\mathfrak{m}$ hadron-hadron channels which open at thresholds, \smash{$E^{(1)}_{\text{thr.}} < E^{(2)}_{\text{thr.}} < \dots < E^{(\mathfrak{m})}_{\text{thr.}}$}. For each hadron-hadron channel $a$, there may be multiple partial-waves under consideration. Suppose channel $a$ has $\mathfrak{n}_a$ partial waves, then $\sum_{a=1}^\mathfrak{m} \mathfrak{n}_a = \mathfrak{n}$, the dimension of the matrix $\bm{D}(E_\cm)$.\footnote{Some partial waves may appear embedded more than once.}

For energies \smash{$E_\cm \geqslant E^{(\mathfrak{m})}_{\text{thr.}}$}, the finite-volume matrix $\bm{\mathcal{M}}(E_\cm,L)$ is hermitian\,\footnote{See Appendix~\ref{AppA}.}, and then since $i\bm{\mathcal{M}}$ is skew-hermitian, it has purely imaginary eigenvalues meaning that $(\bm{1}-i\bm{\mathcal{M}})^{-1}$ exists for all such energies. It immediately follows that $\bm{V}$ is a unitary matrix for all \smash{$E_\cm \geqslant E^{(\mathfrak{m})}_{\text{thr.}}$}. As $\bm{S}$ is also unitary in this energy region by conservation of probability, commonly referred to as {\em unitarity}, $\bm{{D_V}}(E_\cm)$ is diagonalisable with $\mathfrak{n}$ linearly independent orthonormal eigenvectors and bounded eigenvalues of the form
\begin{equation}\label{SecII:Eq:eigs}
\lambda_p(E_\cm)=2\cos \big( \tfrac{1}{2} \theta_p(E_\cm) \big)\, \exp \big(i \tfrac{1}{2}\theta_p(E_\cm) \big) \,,
\end{equation} 
where the impact of unitarity is to ensure that the complex eigenvalues are each controlled by a single real parameter, $\theta_p$.

Since $\bm{{D_V}}(E_\cm)$ is continuous in $E_\cm$, a small perturbation in energy, $E_\cm^{\,\prime} = E_\cm + \Delta E_\cm$, which we will refer to as an \emph{iteration} in energy, results in a small perturbation of the matrix, $\bm{D_V}(E_\cm^{\,\prime}) = \bm{D_V}(E_\cm)+  \bm{\Delta D_V}(E_\cm)$, where $|| \bm{\Delta D_V}(E_\cm)||=O(\Delta E_\cm)$.
It immediately follows from standard matrix perturbation theory that the eigenvalues and eigenvectors at the next iteration, $\lambda_p^\prime \equiv \lambda_p(E_\cm^{\,\prime})$, $\bm{v}^{(p)\prime} \equiv \bm{v}^{(p)}(E_\cm^{\,\prime})$ are approximated in terms of the eigenvalues and eigenvectors at the current iteration, $\lambda_p \equiv \lambda_p(E_\cm)$, $\bm{v}^{(p)} \equiv \bm{v}^{(p)}(E_\cm)$, by

\begin{align*}
\lambda_p^\prime &= \lambda_p\, +\, \bm{v}^{(p)\dagger} \!\cdot\!  \bm{\Delta D_V} \!\cdot\! \bm{v}^{(p)}
+ O\big(|| \bm{\Delta D_V}||^2 \big) 
\\[1.5ex]
\bm{v}^{(p)\prime} &= \bm{v}^{(p)} \,+ \sum_{k\neq p} \frac{\bm{v}^{(k)\dagger} \!\cdot\!  \bm{\Delta D_V}  \!\cdot\! \bm{v}^{(p)}}{\lambda_p -\lambda_k} \, \bm{v}^{(k)} + O\big(|| \bm{\Delta D_V}||^2 \big), 
\end{align*}
so that for the case of non-degenerate eigenvalues, the eigenvectors are indeed small perturbations for $\Delta E_\cm$ sufficiently small. The inner product between eigenvectors on consecutive energy iterations is given by
\begin{align}\label{SecII:Eq:inner_prod}
\bm{v}^{(p)\prime \dagger} \cdot \bm{v}^{(q)} = \delta_{pq} + \Delta_{pq} \, ,
\end{align}
where
\begin{equation*}
\Delta_{pq} = 
\frac{  \bm{v}^{(q)\dagger} \!\cdot\!  \bm{ \Delta D_V} \!\cdot\! \bm{v}^{(p)} }
{\lambda_p - \lambda_q } + O \big(|| \bm{\Delta D_V}||^2 \big) \, ,
\end{equation*}
evaluated at $E_\cm$ will be $\ll 1$ for a sufficiently small choice of $\Delta E_\cm$.
It follows that eigenvectors (and their corresponding eigenvalues) can be matched between energy iterations by examining Eqn.~\ref{SecII:Eq:inner_prod}, with the order of the eigenvectors at $E_\cm^{\, \prime}=E_\cm + \Delta E_\cm$ taken so as to maximise the sum of the norm-squared of the inner products with the eigenvectors at $E_\cm$.

In the case that a number of eigenvalues are degenerate or very nearly degenerate, i.e. when~${\lambda_p -\lambda_k = O\big(||\bm{\Delta D_V }||\big)}$ for at least one $k\neq p$, matching the corresponding eigenvectors between energy iterations is potentially ambiguous. As discussed in Appendix~\ref{AppB}, in these cases it is common to observe an {\em avoided crossing}, where the eigenvalues repel each other rather than cross, but crossings are possible when channels or partial-waves are decoupled. For an energy $E_\cm$ where the eigenvalues are degenerate and cross, the eigenvectors span the corresponding eigenspace and there is an arbitrariness in their direction. At this energy, it is not necessary to match the eigenvectors to the previous iteration, as we are interested in the eigenvalues, which are equal. At the \emph{next} energy iteration, $E_\cm + \Delta E_\cm$, where the eigenvalues are now distinct, rather than matching to the previous set of eigenvectors at $E_\cm$, we can simply match to those at $E_\cm - \Delta E_\cm$.

\begin{figure*}[t]
	\centering
	\includegraphics[trim={0cm 0cm 0cm 0cm},clip,width=1\textwidth]{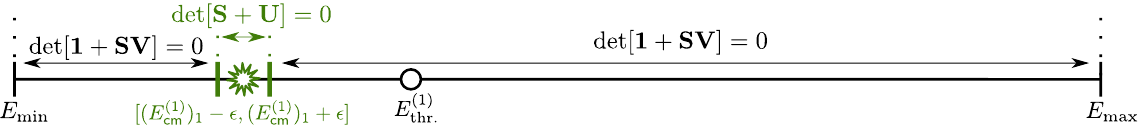}
	\caption{A schematic example of how the various determinant equations are used to find roots of the quantisation condition for a scattering system with a single hadron-hadron channel in an arbitrary number of partial-waves. For a particular energ,y $(E^{(1)}_\cm)_1$, below threshold, $\det[\bm{1}-i\bm{\mathcal{M}}_{11}((E^{(1)}_\cm)_1,L)]=0$, so an energy region is defined in which $\det \big[ \bm{S} + \bm{U} \big]$ is considered rather than $\det \big[ \bm{1} +\bm{S}\cdot \bm{V} \big]$.}
	\label{SecII:Fig:regions}
\end{figure*}

%%%%%%%%%%%%%%%%%%%%%%%%%%%%%%%%%%%%%%%%%%%%%%%%%%%%%%%%%%%%%%%%%%%%%%%%%%%%%%%%%%%%%%%%%%%%%%%%%%
\subsection{With at least one closed channel}\label{SecII:B}
%%%%%%%%%%%%%%%%%%%%%%%%%%%%%%%%%%%%%%%%%%%%%%%%%%%%%%%%%%%%%%%%%%%%%%%%%%%%%%%%%%%%%%%%%%%%%%%%%%

%
One might expect that the finite-volume spectrum at energies below the kinematic threshold for some channel $a$ to be insensitive to amplitudes describing channel $a$ -- after all, there are an infinite number of channels whose thresholds lie higher than any given energy, and we cannot ever possibly know the amplitude in all of them. In fact, the finite-volume spectrum obtained from Eqn.~\ref{Intro:Eq:1+irt} is only sensitive to closed channels in a very small energy region below the threshold, with the effect of the channel reducing exponentially (as $e^{-\kappa L}$ where $\kappa$ is the magnitude of the imaginary scattering momentum). Nevertheless it is important to account for these closed-channel effects in practical calculation, particularly when a resonance appears near a threshold.

At energies below the highest threshold under consideration, \smash{$E_\cm \leqslant E^{(\mathfrak{m})}_{\text{thr.}}$}, the finite-volume matrix $\bm{\mathcal{M}}$ is in general no longer hermitian and subsequently $\bm{V}$ is no longer unitary.  $(\bm{1}-i\bm{\mathcal{M}})$ is not prevented from having zero eigenvalues and correspondingly $\bm{V}$ may diverge at some energies. 

Recalling that $\bm{\mathcal{M}}$ is block-diagonal in channel space, we can solve for roots of $\det[\bm{1}-i\bm{\mathcal{M}}] = 0$ in each channel separately. At decreasing energies below the threshold for channel $a$, $\bm{\mathcal{M}}_{aa}$ tends to $i\bm{1}$ exponentially~\cite{Briceno:2019nns} and so sufficiently far below threshold, $\det[\bm{1}-i\bm{\mathcal{M}}_{aa}]\rightarrow 2^{\mathfrak{n}_a}$. For each channel a root can only be found below threshold, as above threshold we recall $\bm{\mathcal{M}}_{aa}$ is hermitian and no solutions to $\det[\bm{1}-i\bm{\mathcal{M}}_{aa}] = 0$ exist. In order to ensure that zeroes of multiplicity higher than one are found, it is good practice to make use of an eigen-decomposition of $\bm{\mathcal{M}}_{aa}$.

A practical approach to dealing with these isolated energies at which $\bm{V}$ diverges, is to eigen-decompose instead the matrix $\bm{D_U}$ in a small region around the divergent energy, since $\bm{U}$ is typically well defined at those energies. If it should happen by coincidence that $\det[\bm{1}+i\bm{\mathcal{M}}_{aa}]$ has a zero very close to where $\det[\bm{1}-i\bm{\mathcal{M}}_{aa}]$ has a zero, then we could eigen-decompose the original matrix $\bm{D}$. We stress that in all practical applications, this finely tuned scenario seldom occurs, and can be avoided by considering sufficiently small energy regions. Fig.~\ref{SecII:Fig:regions} provides a pictorial representation of the approach.

Finally, there can be divergences in the determinant condition below the {\em lowest} threshold, where no channel is kinematically accessible, when the scattering system features stable bound-states which manifest as simple pole singularities in $\bm{S}(E_\cm)$. The divergence appears at the volume-independent bound-state energy, which differs from a zero corresponding to a finite-volume energy level only by an exponentially small finite-volume correction.

%%%%%%%%%%%%%%%%%%%%%%%%%%%%%%%%%%%%%%%%%%%%%%%%%%%%%%%%%%%%%%%%%%%%%%%%%%%%%%%%%%%%%%%%%%%%%%%%%%%%%%%%%%%%%%
\subsection{At kinematic thresholds}\label{SecII:C}
%%%%%%%%%%%%%%%%%%%%%%%%%%%%%%%%%%%%%%%%%%%%%%%%%%%%%%%%%%%%%%%%%%%%%%%%%%%%%%%%%%%%%%%%%%%%%%%%%%%%%%%%%%%%%%

%
The energy at which a new channel opens, its threshold, requires special consideration when $\bm{D_V}$ or $\bm{D_U}$ are used. If the threshold coincides with a non-interacting energy, such as in $S$-wave scattering in the rest frame, then there is a divergence due to a pole singularity of finite order in $\bm{D}$ when $E_\cm=E_{\text{thr.}}$, which is removed in $\bm{D_V}$ and $\bm{D_U}$. 

Recall that the matrices $\bm{U}$ and $\bm{V}$ are block-diagonal with respect to hadron-hadron scattering channels, being dense in the partial-wave space only. The threshold divergence in $\bm{\mathcal{M}}$ is such that at threshold $\bm{V}$ and $\bm{U}$ are diagonal and equal to $-\bm{1}$.

In general $\bm{S}$ is block-diagonal in partial-wave space, but dense in hadron-hadron channels, however at the opening of a new threshold, say for channel $a$, the vanishing of the phase-space for this channel means that for any \mbox{partial-wave}, $S_{aa}(E_{\text{thr.}}^{(a)})= 1$ and $S_{ab}(E_{\text{thr.}}^{(a)}) = 0$ for $b \neq a$, while elements $S_{bb}(E_{\text{thr.}}^{(a)})$ are unconstrained.  It follows that at each threshold there will be at least one zero eigenvalue of $\bm{D_V}$ or $\bm{D_U}$, but these zeroes should not be mistakenly considered to be actual finite-volume energy levels -- as they appear at exactly the known threshold energies, there should be no risk of this misidentification.

%%% Root finding %%%
\subsection{Summary of the approach}\label{SecII:D}
Standard eigen-decomposition routines can be applied to $\bm{D_V}$, $\bm{D_U}$ or $\bm{D}$ (whichever is appropriate given the discussion in Section.~\ref{SecII:B}) on a discrete sampling of energies, and the eigenvalues $\lambda_p(E_\cm)$ matched between energy iterations using similarity of the corresponding eigenvectors established through magnitude of inner products. Each eigenvalue can then be considered separately, and typically we find that root-finding on any one of these eigenvalues is no more difficult than in the case of elastic scattering. Energy regions in which root-finding is performed are separated by hadron-hadron thresholds and by regions around divergences in $\bm{V}$, but as these regions are disjoint, the complete set of solutions is the combination of the sets of solutions of each eigenvalue in each energy region. It is therefore not necessary to match eigenvalues across boundaries. 

The combination of the sets of solutions in each eigenvalue gives the full set of roots of $\det \big[\bm{D}(E_\cm) \big]=0$, and of course with the eigenvalue roots in hand it is trivial to check that they are in fact roots of $\det \big[\bm{D}(E_\cm) \big]$.

%\pagebreak
\section{Application: Vector-Vector Scattering}\label{SecIII}
\
Vector-vector scattering provides a useful demonstration of the eigen-decomposition approach described in the previous section, as even in the simple case of total momentum zero we can have a situation in which solutions of high multiplicity appear in the non-interacting limit, while in the interacting case the corresponding zeroes may be very closely spaced.

We will illustrate our approach using a simple toy model of vector-vector scattering in which we have a single hadron-hadron channel comprising two identical vector mesons. Considering the finite-volume spectrum in the $[000]\,E^+$ irrep, we note that Bose symmetry constraints limit the number of contributing partial-waves, and we summarize those that can be non-zero for $J \leqslant 4$ in Table~\ref{SecIII:Tab:VV}.

\begin{table}[b]
{\renewcommand{\arraystretch}{1.2}
	\centering
\begin{tabular}{ c | c} 
	\centering
$J^P$ & $\SlJ$ \\
\hline
$2^+$ & $\,\,\prescript{5\!}{}{S}_2,\,\, \prescript{1\!}{}{D}_2,\,\, \prescript{5\!}{}{D}_2,\,\, \color{jlab_gray}{\prescript{5\!}{}{\mathit{G\!}}_2} $  \\
$4^+$ & $ \,\, \prescript{5\!}{}{D}_4,\,\,\color{jlab_gray}{\prescript{1\!}{}{\mathit{G}}_4,\,\,\prescript{5\!}{}{\mathit{G}}_4,\,\,\prescript{5\!}{}{\mathit{I}}_4} $  \\
\end{tabular}
\caption{Subduction of \mbox{partial-waves}, $\SlJ$, for $J\leqslant 4$ into the $E^+$ irrep for two identical vector mesons. The rows denote the \mbox{partial-wave} content for a given $J^P$, with multiple $\SlJ$ entries indicating \mbox{partial-waves} which mix dynamically. \mbox{Partial-waves} with $\ell \geqslant 4 $ are shown in grey. This table is derived from Table~2 of~\cite{Johnson:1982yq}.}
\label{SecIII:Tab:VV}
}
\end{table} 

We engineer a dense spectrum of energy levels by considering relatively weak interactions in all partial-waves, truncating at $\ell=2$. As indicated in Table~\ref{SecIII:Tab:VV}, this results in the contribution of \emph{four} partial waves, $\prescript{5\!}{}{S}_2, \prescript{1\!}{}{D}_2,\prescript{5\!}{}{D}_2$ and $\prescript{5\!}{}{D}_4$. 

The vector particles are taken to have mass $m=0.5$, and we consider a volume $L = 70$. In this case the non-interacting energies and the corresponding multiplicities are presented in Table~\ref{SecIII:Tab:VV_nonint}, which indicates that we expect for a weakly interacting scattering system to find one level close to threshold, three closely-spaced levels somewhat higher in energy and four more levels a little higher still.
\begin{table}[t]
{\renewcommand{\arraystretch}{1.1}
	\centering
\begin{tabular}{ c | c | c} 
	\centering
$\left(\frac{L}{2\pi} \right)^2 \big|\vec{p\,} \big|^2$ & mult.  & $E_{\text{n.i}}$ \\
\hline
0 	& 	1		& $1.00000$ \\
1 	&   3		& $1.01599$ \\
2   &   4       & $1.03172$ 
\end{tabular}
\caption{Non-interacting energies $E_{\text{n.i}}$, for the lowest allowed back-to-back momenta, along with the corresponding multiplicities that appear in $[000]E^+$ for a pair of vector particles of mass $m = 0.5$ in a volume $L=70$.}
\label{SecIII:Tab:VV_nonint}
}
\end{table}

%%%%%%%%%%%%%%%%%%%%%%%%%%%%%%%%%%%%%%%%%%%%%%%%%%%%%%%%

\begin{figure}[b]
	\centering
	\includegraphics[trim={0cm 0cm 0cm 0cm},clip,width=0.5\textwidth]{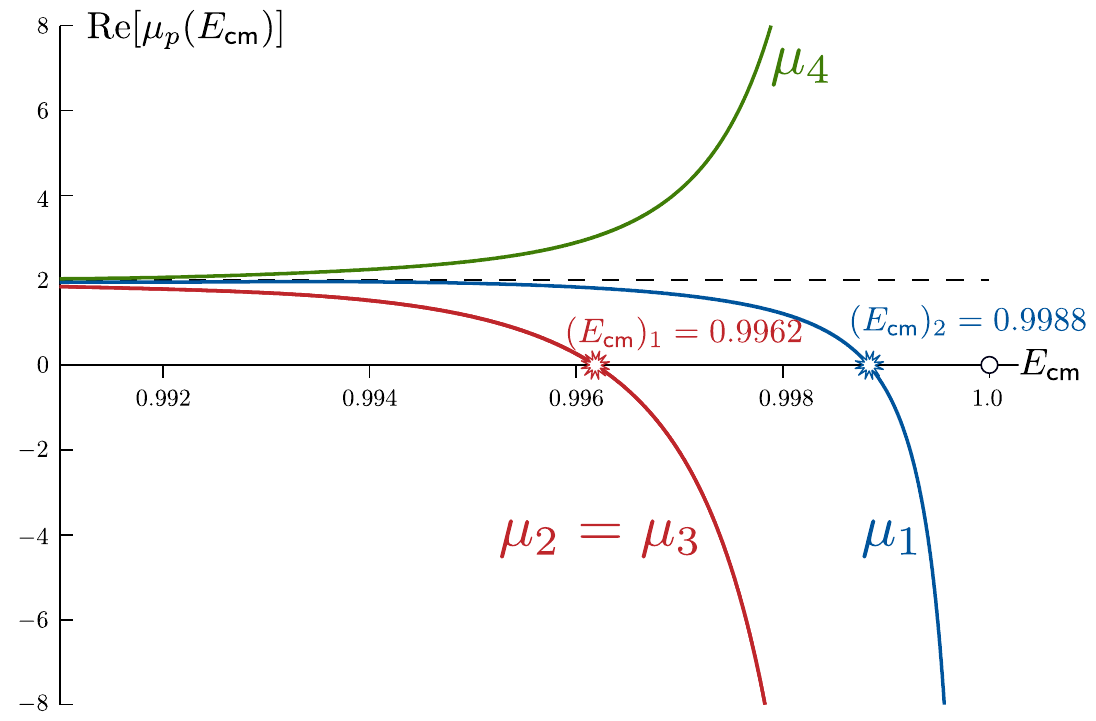}
	\caption{Eigenvalues of $\bm{1}-i\bm{\mathcal{M}}(E_\cm,L)$ for $E_\cm \leqslant E^{(1)}_{\text{thr.}}$. The imaginary part of each eigenvalue is identically zero below threshold.}
	\label{SecIII:Fig:Ecma}
\end{figure}

Since there may be finite-volume energy levels below the opening of our threshold, we should first establish if there are energies around which we should use $\bm{D_U}$ rather than our preferred choice $\bm{D_V}$. This can be done without knowledge of the scattering amplitude, by looking for roots of $\det[\bm{1}-i\bm{\mathcal{M}}]=0$ below threshold. Fig.~\ref{SecIII:Fig:Ecma} shows the eigenvalues of $\bm{1}-i\bm{\mathcal{M}}$ and their roots. Note that each eigenvalue diverges at the threshold, and asymptotically approaches a value of $2$ (as $\bm{\mathcal{M}}\rightarrow i\bm{1}$) as energies go further below threshold. Two roots are found: a double root at $(E_\cm)_1=0.9962$ and a single root at $(E_\cm)_2=0.9988$. A small energy region\,\footnote{For the $m,L$ combination used here, an energy region $1.0\!\times\! 10^{-5}$ either side of the root is quite reasonable.} around each of these values is chosen to be treated with $\bm{D_U}$ in place of $\bm{D_V}$.

To describe scattering in this system of four partial-waves we use a $K$-matrix parameterisation which automatically implements the required unitarity of the $S$-matrix. The $K$-matrix, $\bm{K}(s)$, where $s = E_\mathsf{cm}^2$, is a real symmetric matrix, related to the $t$-matrix via
\begin{align}\label{Eq:Kmat}
\big[\bm{t}^{-1}(s)\big]_{S \ell J, S' \ell' J} = 
\frac{1}{(2k_\cm)^\ell}
& \big[\bm{K}^{-1}(s)\big]_{S \ell J, S' \ell' J}
\frac{1}{(2k_\cm)^{\ell'}} \nonumber \\
+& \delta_{S S'} \delta_{\ell \ell'} \, I(s).
\end{align}
Unitarity is guaranteed if $\text{Im}\,I(s)=-\rho(s)$ for energies above threshold and $\text{Im}\,I(s)=0$ below. We make use of the Chew-Mandelstam prescription~\cite{Chew:1960iv} which defines $\text{Re}\,I(s)$ through a dispersive integral featuring $\rho(s)$, and choose to subtract so that $\text{Re}\,I=0$ at threshold. 

\begin{figure}[t]
	\centering
	\vspace{1cm}
	\includegraphics[trim={0cm 0cm 0cm 0cm},clip,width=0.5\textwidth]{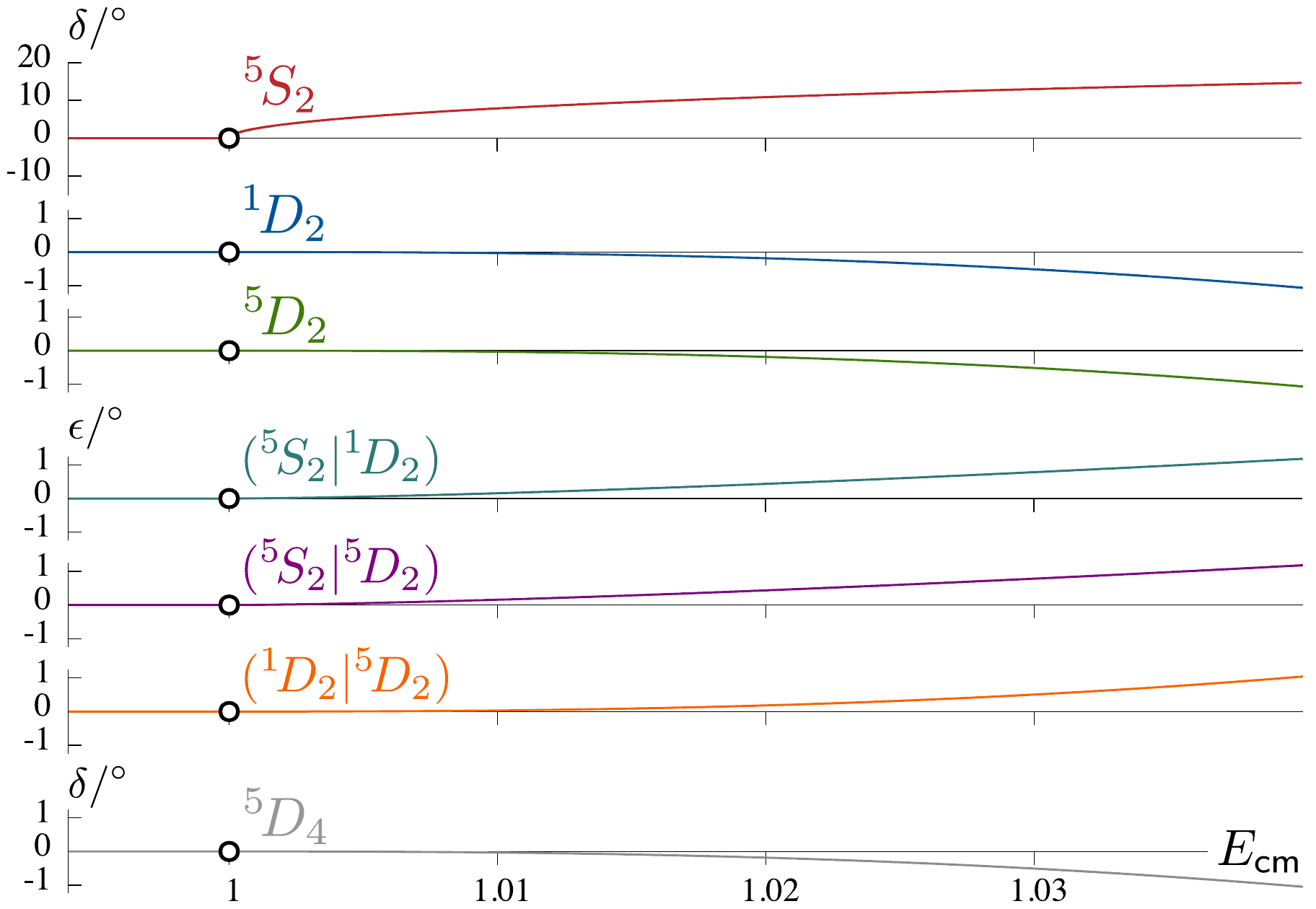}
	\caption{Phase-shifts and mixing-angles for the toy model amplitude given in Eq.~\ref{SecIIIA:Eq:Kmat} following the $n$-channel Stapp parameterisation described in Ref.~\cite{Woss:2019hse}.}
	\label{SecIII:Fig:phases_angles}
\end{figure}

\pagebreak
To implement weak scattering we use a constant \mbox{$K$-matrix} parameterisation featuring small values,
\begin{equation}\label{SecIIIA:Eq:Kmat}
\bm{K}(s) =
\begin{blockarray}{ccccc}
\fiveStwo & \oneDtwo & \fiveDtwo& \fiveDfour\; &  \\[0.6ex]
\begin{block}{[cccc]l}
1 & 1 & 1 &  & \;\;\;\fiveStwo \\  
1 & -10 & 10 & & \;\;\;\oneDtwo \\  
1 & 10 &-10 & & \;\;\;\fiveDtwo  \\  
&  &  & -10 & \;\;\;\fiveDfour \\
\end{block}
\end{blockarray} \, ,
\end{equation}
where the block diagonalization in $J$ is made explicit. The somewhat larger values in the $D$-wave amplitudes are compensated by the threshold factors $k_\cm^\ell$ in Eq.~\ref{Eq:Kmat}. The resulting phase-shifts and mixing-angles for this toy amplitude are presented in Fig.~\ref{SecIII:Fig:phases_angles} using the $n$-channel generalization of the Stapp parameterisation described in Ref.~\cite{Woss:2019hse}. 
Note that all amplitudes are modest by design, with the $S$-wave an order of magnitude larger than any of the $D$-wave amplitudes. 

\begin{figure*}[tbh]
	\centering
	\vspace{1cm}
	\includegraphics[trim={0cm 0cm 0cm 0cm},clip,width=1\textwidth]{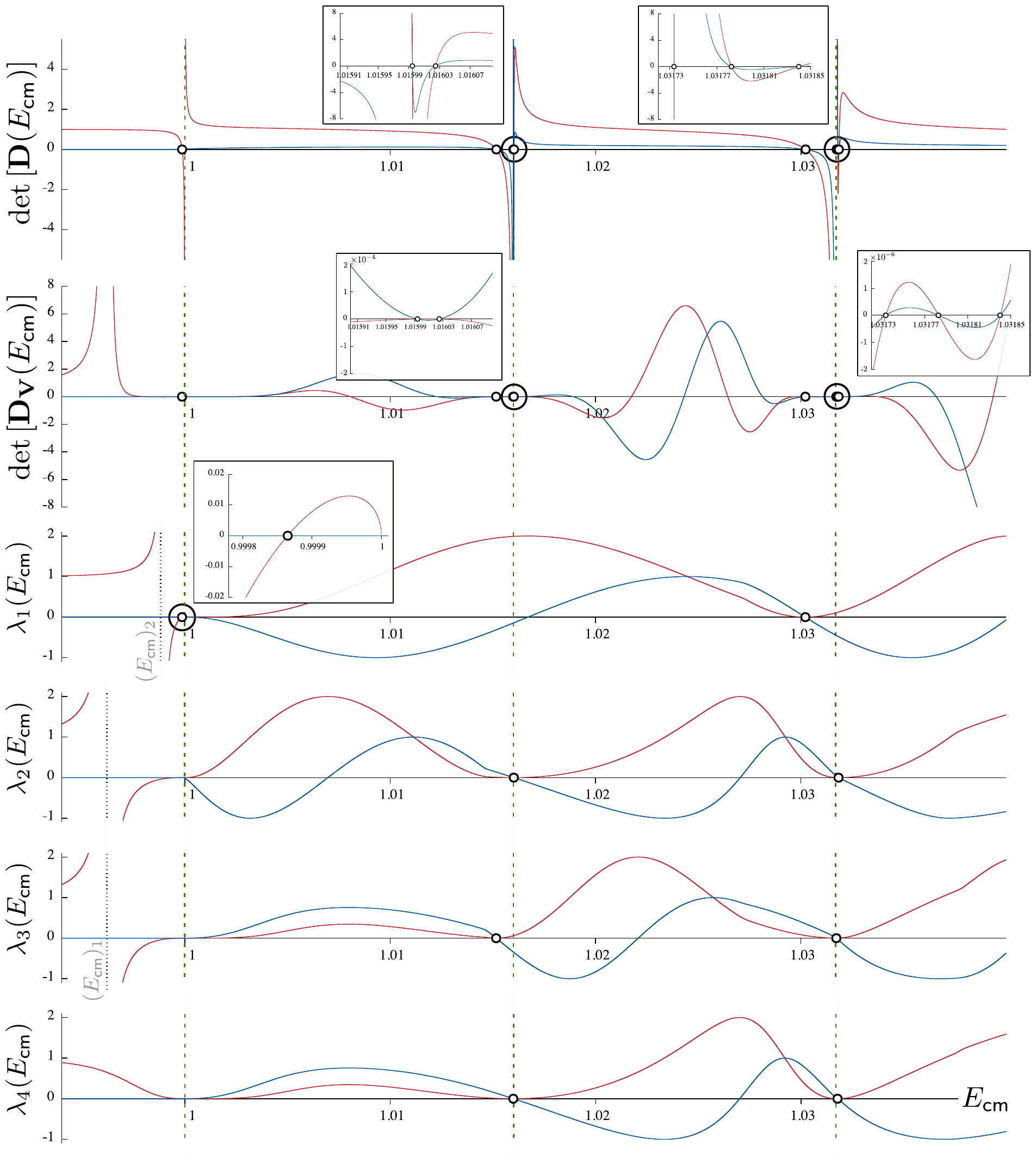}
	\caption{Quantization conditions for the toy model of vector-vector scattering described in the text. Kinematic threshold at $E_\cm = 1.0$ and non-interacting energies indicated by the vertical green dashed lines. Red, blue curves indicate real, imaginary parts respectively. Top panel: $\det \!\big[ \bm{D} ]$ as defined in Eqn.~\ref{Intro:Eq:1+irt}. Second panel: $\det \!\big[ \bm{D_V} ]$ as defined in Eqn.~\ref{SecII:Eq:1+sv}. Bottom four panels: Eigenvalues of $\det \!\big[ \bm{D_V} ]$ -- divergences below threshold at energies presented in Fig.~\ref{SecIII:Fig:Ecma}.
	} 
	\label{SecIII:Fig:Eigs}
\end{figure*}

Following the procedure set out in Sec.~\ref{SecII}, we seek to find solutions of $\det[\bm{D}]=0$. The top panel of Fig.~\ref{SecIII:Fig:Eigs} shows $\det[\bm{D}]$ as defined in Eqn.~\ref{Intro:Eq:1+irt} as a function of energy, where divergences at each of the non-interacting energies can be clearly seen, along with the near-degenerate roots located in very close proximity -- as anticipated from our weakly interacting amplitudes. Finding the zeroes directly from this determinant requires ultra-fine sampling in energy, and roots are easily missed in automated root-finding algorithms. Furthermore, the toy amplitudes could be easily modified to make the near degenerate roots, shown in the zoomed regions, exactly degenerate, in which case no amount of improvement in sampling resolution would find the appropriate number of roots.

The second panel in Fig.~\ref{SecIII:Fig:Eigs} shows $\det[\bm{D_V}]$ as a function of energy. As discussed in Sec.~\ref{SecII}, the divergences that appear at the non-interacting energies have been removed and above threshold the determinant is bounded. In a neighbourhood around the near degenerate roots, the determinant is more clearly seen to be locally quadratic and cubic respectively.

The remaining lower panels in Fig.~\ref{SecIII:Fig:Eigs} show the four eigenvalues of  $\det[\bm{D_V}]$ over the considered energy range, about which we make a number of observations. Firstly, the real part of each eigenvalue is in the interval $[0,2]$ above threshold and the imaginary component in $[-1,1]$ as proven in Sec.~\ref{SecII:A}. Below threshold, the eigenvalues indeed diverge at energies where $\bm{V}$ is singular as expected -- see Sec.~\ref{SecII:B}. The multiplicities of these divergences also agree, for example as shown in Fig.~\ref{SecIII:Fig:Ecma} the determinant $\det[\bm{1}-i\bm{\mathcal{M}}]$ has a double root at $(E_\cm)_1$, which manifests in the eigenvalues $\lambda_2$ and $\lambda_3$, and a single root at $(E_\cm)_2$, which appears in $\lambda_1$. At threshold, each eigenvalue is identically zero as proven in Sec.~\ref{SecII:C} and these threshold zeroes are removed from the set of solutions. It is clear from counting the number of roots that we have agreement with the multiplicities of the non-interacting energies, shown in Table~\ref{SecIII:Tab:VV_nonint}, as we expect for weak interactions. Most notably, finding roots in each eigenvalue is no more difficult than solving for roots in elastic scattering and this is illustrated by the simplicity of the zero crossings in the eigenvalues. The found roots are recorded in Table~\ref{SecIII:Tab:roots}.

%%%%%
\begin{table}[h]
	{\renewcommand{\arraystretch}{1.2}
		\begin{tabularx}{0.3\textwidth}{c @{\extracolsep{\fill}} ccc}
			$\lambda_1$ & $\lambda_2$ & $\lambda_3$ & $\lambda_4$ \\
			\hline
			0.999865 & & & \\
			 & 1.01602 & 1.01516 & 1.01599 \\
			1.03022 & 1.03184 & 1.03173 & 1.03178 \\
			%%%%%
			%%%%%
		\end{tabularx}
	}
	\caption{Solutions of $\det[\bm{D}]=\det[\bm{D_V}]=0$ and eigenvalues, $\lambda_p$, of $\bm{D_V}$ in which they appear. Values grouped according to the non-interacting multiplicities provided in Table~\ref{SecIII:Tab:VV_nonint}. }
	\label{SecIII:Tab:roots}
\end{table}

\section{Application: $1^+$, $1^-$ Scattering}\label{SecIV}

Previous lattice QCD calculations have determined the lowest lying isospin=1 $J^{PC}=1^{--},1^{+-}$ vector~\cite{Dudek:2012xn} and axial-vector~\cite{Woss:2019hse} resonances, the $\rho$ and $b_1$, in studies on three lattice volumes\,\footnote{Details of the lattices, and the correlator construction technology can be found in Refs.~\cite{Lin:2008pr, Peardon:2009gh}. } with a pion mass  $\sim 391 \text{ MeV}$.
In the determination of the $b_1$ resonance, energy levels used to constrain the scattering amplitudes were obtained in the $[000]\,T_1^+$ and $(\vec{P} \neq \vec{0})\,A_2$ irreps, as these irreps receive no contribution from $J^P=1^-$ scattering amplitudes. This avoids the complication of disentangling the low-lying, resonating $1^{-}$ channels that mix due to the reduced symmetry of the finite-volume. The eigen-decomposition method we have introduced in this manuscript enables us to lift such practical restrictions and perform a combined scattering analysis of the $1^{--}$ and $1^{+-}$ sectors. Using a much larger set of irreps, including now those in which both parities feature, allows us to significantly increase the number of energy levels constraining the scattering amplitudes, with a total of 144 energy levels being available.

Several hadron-hadron channels in various partial-waves feature in these scattering systems. For the $J^P=1^-$ sector, in addition to $\pi\pi\{\onePone\}$ considered in Ref.~\cite{Dudek:2012xn} (which restricted itself to elastic energies), we include $K\overline{K}\{\onePone\}$ and $\pi\omega\{\threePone\}$ which are kinematically open in the energy region where where the $b_1$ resonance appears. For $J^P=1^+$, we consider $\pi\omega\{\threeSone\}$, $\pi\omega\{\threeDone\}$ and $\pi\phi\{\threeSone\}$ as in Ref.~\cite{Woss:2019hse}. As discussed in great detail in Ref.~\cite{Woss:2019hse}, three-body thresholds, $\pi\pi\eta$ and $\pi K\overline{K}$, open below the $4\pi$-threshold in the region where the $b_1$ resonates and we ensure that in each irrep, we keep below the lowest $E^{2+1}_{\text{n.i}}$ non-interacting energy~\footnote{As introduced in  Ref.~\cite{Woss:2019hse}, $E^{2+1}_{\text{n.i}}$ non-interacting energies are finite-volume energies calculated in a two-meson subsystem that are then combined with the third meson, where no interactions are assumed between the two-meson subsystem and third meson.}.

Other partial-waves also feature due to the reduced symmetry of the finite-volume. For $\pi\pi$ and $K\overline{K}$ with isospin=1, a $\oneFthree$ wave can appear (see Table~III of Ref.~\cite{Dudek:2012xn}) but will be suppressed by the high value of $\ell$. For vector-pseudoscalar scattering, the nonzero intrinsic spin gives rise to a triplet of partial-waves for each $\ell \geqslant 1$, e.g.~$\threePzero$, $\threePone$ and $\threePtwo$ for $\ell = 1$ (see Table~1 of Ref.~\cite{Woss:2018irj} for irreps at rest, and Tables~5-7 for irreps in flight). Following the analysis in Ref.~\cite{Woss:2019hse}, we include only $\pi \omega \{\threePone\}$, with $J^P=1^-$ which has not been previously shown to be compatible with zero. The resulting set of partial waves to be considered is presented in Table~\ref{b1pw}.

\begin{table}[h]
{\renewcommand{\arraystretch}{1.2}
	\centering
\begin{tabular}{ r | ccc} 
	\centering

$1^-$ & $\pi\pi \{ \onePone \}$	& $K\overline{K}\{ \onePone \}$ 	& $\pi \omega \{ \threePone \}$ \\[2ex]
$1^+$ & $\pi\omega \{ \threeSone \}$	& $\pi\omega \{ \threeDone \}$ 	& $\pi\phi \{ \threeSone \}$
\end{tabular}
\caption{Partial-wave basis to describe finite-volume spectrum.}
\label{b1pw}
}
\end{table} 

As in the case of the toy amplitudes in Sec.~\ref{SecIII}, singularities of $\bm{V}$ must be determined in order to solve the appropriate form of quantisation condition in each energy region. For reference, tables~\ref{AppC:Tab:rho_irreps} --~\ref{AppC:Tab:rho_b1_irreps} in Appendix~\ref{AppC} report the singularities of $\bm{V}$ on each volume and lattice irrep. 

In this case we are seeking to describe an explicit spectrum of lattice QCD energy levels (with uncertainties and correlations) using a parameterised amplitude in a $\chi^2$ minimization\,\footnote{The $\chi^2$ is defined explicitly in Eqn.~9 of Ref.~\cite{Dudek:2012xn}.}. In the approach followed by \emph{hadspec}, typically a range of parameterisations would be considered, but in this case, where we seek only to demonstrate the effectiveness of the eigen-decomposition approach, we use just one\,\footnote{and in addition we do not consider the effects of varying the stable hadron masses, or the anisotropies within their uncertainties.}, a \textit{K}-matrix parameterisation of the `pole plus constant' form for each $J^P$,
\begin{align}\label{SecIV:Eq:Kmat}
\bm{K}(s) &=  \begin{bmatrix} \bm{K}_{1^-}(s) & \bm{0} \\ \bm{0} & \bm{K}_{1^+}(s) \end{bmatrix}  \, , \nonumber \\[1.5ex]
\bm{K}_{1^-}(s) &= \frac{1}{m_-^2 - s}
\begin{bmatrix}
g^{\,2}_{\pi\pi\{\onePone\}} & 0 & 0 \\
0 							 & 0 & 0 \\
0 							 & 0 & 0
\end{bmatrix} \nonumber \\
&\quad\quad+\;\;
\begin{bmatrix}
\gamma_{\pi\pi\{\onePone\}} & 0 								 & 0  \\
0							& \gamma_{K\overline{K}\{\onePone\}} & 0 \\ 
0 							& 0 								 & \gamma_{\pi\omega\{\threePone\}}
\end{bmatrix} \, , \nonumber \\[1.5ex]
%%%%%%%%%%%%%%%%%%%%%%%%%%%%%%%%%%%%%%%%%%%%%%%%%%
\bm{K}_{1^+}(s) &= \frac{1}{m_+^2 - s}
\begin{bmatrix}
g^{\,2}_{\pi\omega\{\!\threeSone\!\}} 						& g_{\pi\omega\{\!\threeSone \!\}}g_{\pi\omega\{\!\threeDone\!\}}   & 0 \\
g_{\pi\omega\{\!\threeSone \!\}}g_{\pi\omega\{ \! \threeDone \!\}} & g^{\, 2}_{\pi\omega\{ \!\threeDone \!\}}  					   & 0 \\
0 & 0 & 0 
\end{bmatrix}\nonumber \\
&\quad\quad+\;\;
\begin{bmatrix}
\gamma_{\pi\omega\{\threeSone\}} & 0 & 0 \\
0								 & 0 & 0 \\
0 								 & 0 & \gamma_{\pi\phi\{\threeSone\}}  
\end{bmatrix}.
\end{align}
We use the Chew-Mandelstam prescription for $I(s)$, choosing to subtract at the pole masses so that $\text{Re}\, I_{aa}(m_-^2)=0$ where $a$ runs over channels with $J^P=1^-$ and similarly $\text{Re}\, I_{aa}(m_+^2)=0$ for channels with $J^P=1^+$. The use of only a single pole in each $J^P$ is motivated by the expectation of a narrow $\rho$ resonance, a narrow $b_1$ resonance, and no other resonances until much higher energies~\cite{Dudek:2013yja, Dudek:2010wm}.

The best description of the lattice spectra in irreps $[000]\,T_1^\pm$, $[001]\, (A_1, A_2, E_2)$, $[011]\,(A_1, A_2, B_1, B_2)$, $[111]\, (A_1, A_2, E_2)$ and $[002]\, (A_1, A_2)$, varying the ten parameters in Eqn.~\ref{SecIV:Eq:Kmat}, is found with a quite reasonable $\chi^2 / N_{\text{dof}} = 229. / (144 - 10) = 1.71$, after $\sim 400$ iterations of the MINUIT minimisation routine~\cite{James:1975dr}.

\begin{figure}
	\centering
	\vspace{1cm}
	\includegraphics[trim={0cm 0cm 0cm 0cm},clip,width=0.5\textwidth]{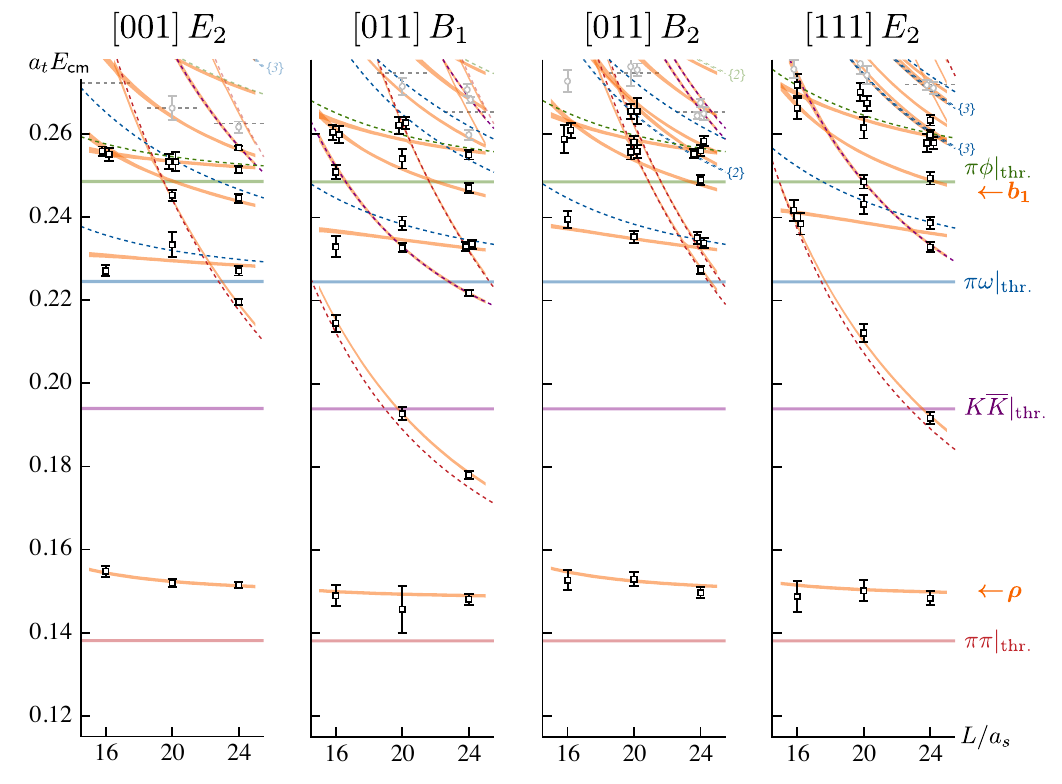}
	\caption{For those irreps which feature both $J^P=1^+$ and $1^-$, the lattice QCD energy spectrum (black points, or grey if points not used in the fit), non-interacting curves (dashed lines) and in orange the finite-volume spectrum corresponding to the amplitude in Eqn.~\ref{SecIV:Eq:Kmat} using the best-fit parameters, including the propagation of their correlated errors.}
	\label{SecIV:Fig:orange_curves}
\end{figure}

Fig.~\ref{SecIV:Fig:orange_curves} shows as orange curves the finite-volume spectrum corresponding to the best-fit amplitude of Eqn.~\ref{SecIV:Eq:Kmat} for those irreps in which both $J^P=1^-$ and $1^+$ are subduced. Also shown are the relevant non-interacting energy curves and the lattice QCD energy levels which were fitted to obtain the best-fit parameter values.

\begin{figure*}[tbh]
	\centering
	\vspace{1cm}
	\includegraphics[trim={0cm 0cm 0cm 0cm},clip,width=1\textwidth]{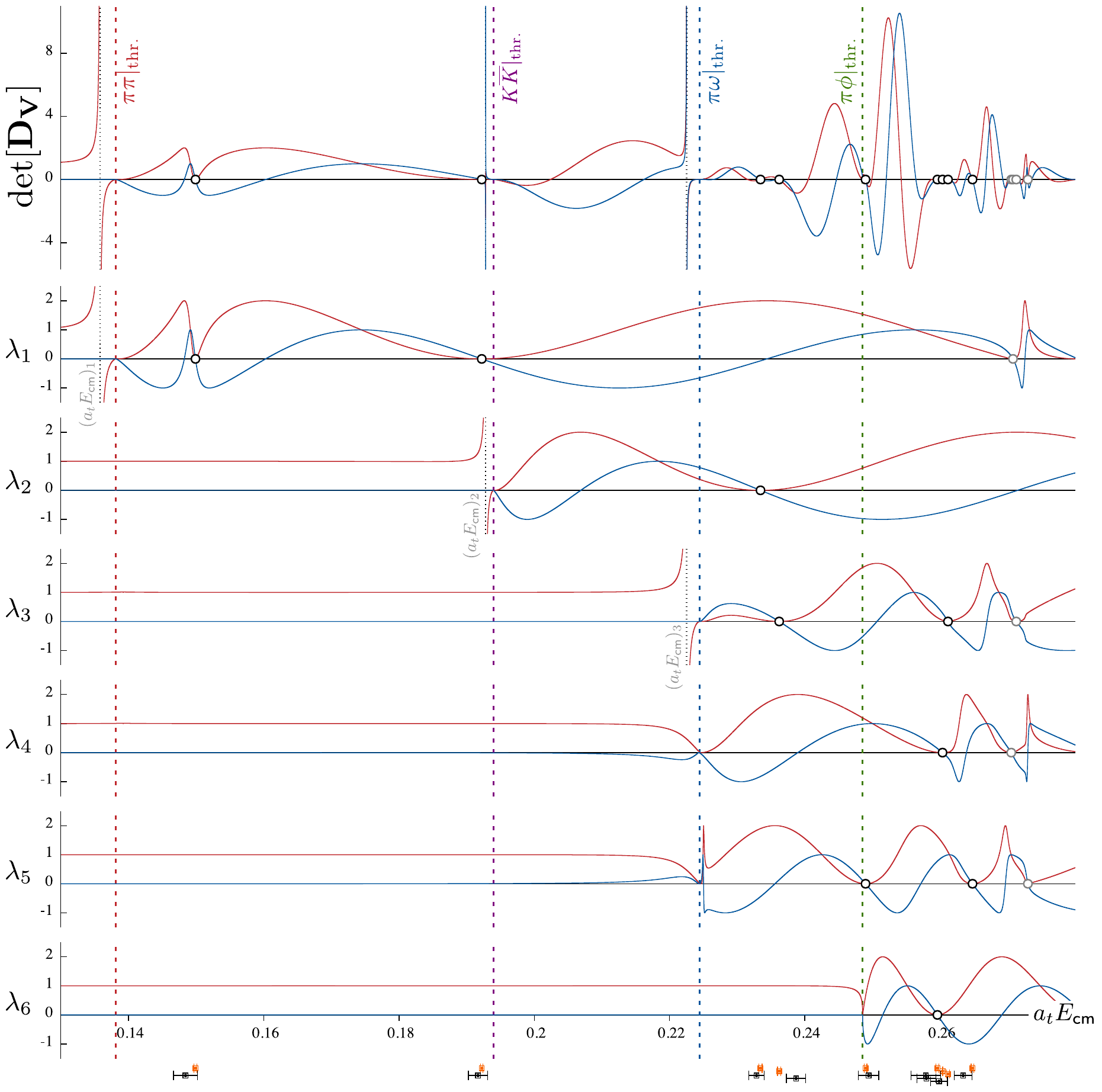}
	\caption{Quantization condition for the $[111]\,E_2$ irrep on the $L/a_s=24$ lattice described in the text. Kinematic thresholds for $\pi\pi$, $K\overline{K}$, $\pi\omega$ and $\pi\phi$ at $a_tE_\cm = 0.1381, 0.1940, 0.2245$ and $0.2486$ respectively, indicated by the coloured vertical dashed lines. Vertical dotted grey lines correspond to singularities of $\bm{V}$ recorded in Table~\ref{AppC:Tab:rho_b1_irreps} in Appendix~\ref{AppC}. Red, blue curves indicate real, imaginary parts respectively. Top panel: $\det \!\big[ \bm{D_V} ]$ as defined in Eqn.~\ref{SecII:Eq:1+sv}. Bottom six panels: Eigenvalues of $\det \!\big[ \bm{D_V} ]$. Black points at the foot of the figure denote the lattice computed energies transcribed from Figure~\ref{SecIV:Fig:orange_curves}. Orange points are the roots of $\det \!\big[ \bm{D_V} ]$ with statistical uncertainties reflecting the uncertainties on the scattering parameters.
	} 
	\label{SecIV:Fig:Eigs}
\end{figure*}

In Fig.~\ref{SecIV:Fig:Eigs} we illustrate the solution of the quantization condition in the case of the $[111]\,E_2$ irrep on the $L/a_s = 24$ lattice using the central values of the best fit parameters. A single embedding of each partial-wave in each hadron-hadron channel recorded in Table~\ref{b1pw} features in this irrep, and thus $\bm{D_V}$ is a $6\times 6$ matrix. We make a number of observations. First, as demonstrated in the toy model example in Sec~\ref{SecIII}, the multiplicities of the singularities of $\bm{V}$ (see Table~\ref{AppC:Tab:rho_b1_irreps}) are manifested in the eigenvalue divergences, i.e. three singularities with multiplicity one. The determinant of $\bm{D_V}$ is indeed bound above the $\pi\phi$ threshold and moreover, as no singularity of $\bm{D_V}$ appears above $\pi\omega$ threshold in this case, $\bm{D_V}$ is also bound between $\pi\omega$ and $\pi\phi$ threshold. For this particular parameterisation, the $S$-matrix is decoupled in hadron channel (see Eq.~\ref{SecIV:Eq:Kmat}). As $\bm{V}$ is diagonal in hadron channel, $\bm{D_V}$ is block-diagonal and the eigenvectors are therefore orthogonal in channel space. This is manifest in the corresponding eigenvalues and we can make the following associations: $\lambda_1$ corresponds to $\pi\pi\{\onePone\}$, $\lambda_2$ to $K\overline{K}\{\onePone\}$, $\lambda_3,\lambda_4,\lambda_5$ to $\pi\omega\{\threeSone,\threePone,\threeDone\}$ and $\lambda_6$ to $\pi\phi\{\threeSone\}$. Of course this is not a general feature of the eigen-decomposition method as the $S$-matrix is generally dense in channel space. The roots and the corresponding eigenvalues of $\bm{D_V}$ in which they appear are recorded in Table~\ref{SecIV:Tab:roots}. 

%%%%%
\begin{table}[h]
\hspace{-5mm}
%	\setlength{\tabcolsep}{2pt} 
%	\small
	{\renewcommand{\arraystretch}{1.2}
		\begin{tabularx}{0.5\textwidth}{c @{\extracolsep{\fill}} ccccc}
			$\lambda_1$ & $\lambda_2$ & $\lambda_3$ & $\lambda_4$ & $\lambda_5$ & $\lambda_6$ \\
			\hline
			0.149900 & 0.233452 & 0.236217 & 0.260367 & 0.248984 & 0.259602\\
			0.192234 & & 0.261172 & \color{jlab_gray}{0.270523} & 0.264797 &\\
			\color{jlab_gray}{0.270783} & & \color{jlab_gray}{0.271257} & \color{jlab_gray}{0.272992} & &\\
			%%%%%
			%%%%%
		\end{tabularx}
	}
	\caption{Roots of $\det[\bm{D}]=\det[\bm{D_V}]=0$ in the $[111]E_2$ irrep on the $L/a_s=24$ lattice for the amplitude in Eqn.~\ref{SecIV:Eq:Kmat} listed by the eigenvalue of $\bm{D_V}$ in which they appear. Grey entries denote roots that were found at energies higher than those we are considering.}
	\label{SecIV:Tab:roots}
\end{table}

For this parameterisation, we find the $\pi\pi\{\onePone\}$ amplitude appears to agree very well with the amplitudes determined in the previous calculation~\cite{Dudek:2012xn}, with the energy dependence above the resonance being quite modest. The amplitude features a complex conjugate pair of resonance poles, which we interpret as the $\rho$, at,
\begin{equation}\label{SecIV:Eq:rho_pole}
a_tE_\cm = 0.1493(2) \pm \tfrac{i}{2} 0.00183(6)\, ,
\end{equation}
which is in quite reasonable agreement with the value found previously using a more limited set of irreps. The $K\overline{K}\{\onePone\}$ and $\pi\omega\{\threePone\}$ amplitudes were found to be weak over the entire energy range considered, with phase-shifts not exceeding $1^\circ$ and $17^\circ$ respectively, as anticipated owing to the absence of any $1^-$ resonances between the $K\overline{K}$ and $4\pi$ thresholds at this pion mass. 

Similarly, the $J^P=1^+$ amplitudes were found to agree very well with those found in the previous calculation~\cite{Woss:2019hse}. The amplitude features a complex conjugate pair of resonance poles, which we interpret as the $b_1$, at,
\begin{equation}\label{SecIV:Eq:b1_pole}
a_tE_\cm = 0.2454(9) \pm \tfrac{i}{2} 0.0199(12) .
\end{equation}
in close agreement with the pole determined in Ref.~\cite{Woss:2019hse}.

\section{Summary}\label{SecV}

We have presented an approach for efficiently solving the L\"uscher quantization condition which determines the discrete spectrum of states in a finite-volume given a scattering system. The technique makes use of the eigenvalue decomposition of the matrix appearing under the determinant in the quantization condition, ultimately reducing the complexity of the numerical problem for coupled-channels and coupled partial-waves down to one of root-finding in a number of independent continuous functions which typically each feature far less rapid variation than does the determinant itself.

Two applications were given for illustration. In the first, a toy model of weak scattering in a system of two vector mesons shows the ability of the approach to determine the very dense set of zeroes lying close to non-interacting levels of high multiplicity. In the second, a set of lattice QCD calculated energy levels on three volumes were used to constrain coupled-channel scattering with $J^P = 1^-$ and $1^+$. This provides a stress-test of the approach, as the solving of the quantization condition needs be successful on multiple irreps in multiple volumes, at many hundreds of iterations of the scattering amplitude parameter values, varied under a $\chi^2$ minimization controlled by a minimization routine. The approach was found to be both reliable and computationally fast.

Contemporary examples where application of this approach may prove to make possible previously impractical calculations include the determination of the resonance spectrum in charmonium where a decay product vector meson $D^*$ is well approximated as a stable hadron, and where the $D\bar{D}$, $D^* \bar{D}$ and $D^* \bar{D}^*$ thresholds lie very close to each other. In-flight irreps will receive contributions from many partial-waves, and there is the expectation of resonances in all of $J^{PC}= 1^{--}, 2^{--}, 3^{--}$ with rather similar masses. Nucleon resonances provide another application where a relatively large number of scattering channels and partial-waves contribute in the irreps of interest. We expect our approach to significantly simplify the finding of finite-volume energy levels in systems like these, and hence to make possible the determination of larger parts of the resonance content of QCD.

\vspace{5mm}
\acknowledgments
{
We thank our colleagues within the Hadron Spectrum Collaboration.
AJW is supported by the U.K. Science and Technology Facilities Council (STFC) [grant number ST/P000681/1]. 
JJD acknowledges support from the U.S. Department of Energy contract DE-SC0018416, and contract DE-AC05-06OR23177, under which Jefferson Science Associates, LLC, manages and operates Jefferson Lab. 
DJW acknowledges support from a Royal Society University Research Fellowship and partial support from the U.K. Science and Technology Facilities Council (STFC) [grant number ST/P000681/1].

The software codes
{\tt Chroma}~\cite{Edwards:2004sx} and {\tt QUDA}~\cite{Clark:2009wm,Babich:2010mu} were used. 
The authors acknowledge support from the U.S. Department of Energy, Office of Science, Office of Advanced Scientific Computing Research and Office of Nuclear Physics, Scientific Discovery through Advanced Computing (SciDAC) program.
Also acknowledged is support from the U.S. Department of Energy Exascale Computing Project.
 
This work was performed using the Cambridge Service for Data Driven Discovery (CSD3), part of which is operated by the University of Cambridge Research Computing on behalf of the STFC DiRAC HPC Facility (www.dirac.ac.uk). The DiRAC component of CSD3 was funded by BEIS capital funding via STFC capital grants ST/P002307/1 and ST/R002452/1 and STFC operations grant ST/R00689X/1. DiRAC is part of the National e-Infrastructure.
This work was performed using the Darwin Supercomputer of the University of Cambridge High Performance Computing Service (www.hpc.cam.ac.uk), provided by Dell Inc. using Strategic Research Infrastructure Funding from the Higher Education Funding Council for England and funding from the Science and Technology Facilities Council.
This work was also performed on clusters at Jefferson Lab under the USQCD Collaboration and the LQCD ARRA Project.
This research was supported in part under an ALCC award, and used resources of the Oak Ridge Leadership Computing Facility at the Oak Ridge National Laboratory, which is supported by the Office of Science of the U.S. Department of Energy under Contract No. DE-AC05-00OR22725.
This research used resources of the National Energy Research Scientific Computing Center (NERSC), a DOE Office of Science User Facility supported by the Office of Science of the U.S. Department of Energy under Contract No. DE-AC02-05CH11231.
The authors acknowledge the Texas Advanced Computing Center (TACC) at The University of Texas at Austin for providing HPC resources.
Gauge configurations were generated using resources awarded from the U.S. Department of Energy INCITE program at the Oak Ridge Leadership Computing Facility.
}

%%%%%%%%%%%%%%%%%%%%%%%%%%%%%%%%%%%%%%%%%%%%%%%%%%%%%%%%%%%%%%%%%%%%%%%%%%%%%%%%%
\appendix

\vspace{5mm}
\section{Hermiticity of \texorpdfstring{$\bm{\mathcal{M}}$}{$\mathcal{M}$} above threshold}\label{AppA}

Bounding of the eigenvalues and orthogonality of the eigenvectors of $\bm{D_V}$ followed from hermiticity of $\bm{\mathcal{M}}$ above threshold. Here we show this hermiticity, starting with the general expression for matrix elements ${\mathcal{M}}$ in the basis of intrinsic spin $S$, orbital angular momentum $\ell$, total angular momentum $J$, azimuthal component of total angular momentum $m$ and channel index $a$,
\begin{widetext}
\begin{align}\label{AppA:Eq:M}
\mathcal{M}_{\ell S J m a,\,  \ell' S' J' m' b} = \delta_{ab} \, \delta_{SS'}\sum_{\substack{m_\ell, m_\ell'\\ m_S}} \langle &\ell m_\ell; S m_S | J m\rangle \,\langle \ell' m_\ell'; S m_S | J' m'\rangle \nonumber \\
& \times \sum_{\bar{\ell}, \bar{m}_\ell}  \frac{(4\pi)^{3/2}}{k_\mathsf{cm}^{\bar{\ell} +1} } \, c^{\vec{n}}_{\bar{\ell}, \bar{m}_\ell}(k_\mathsf{cm}^2; L) \, \int\!\! d\Omega \; Y_{\ell m_\ell}^*  Y_{\bar{\ell} \bar{m}_\ell}^* Y^{}_{\ell' m_\ell'} \, .
\end{align}
The spin-orbit coupling is expressed through the real $\text{SU}(2)$ Clebsch-Gordan coefficients $\langle \ell m_\ell; S m_S | J m\rangle$ and the volume dependence is encoded in the functions $c^{\vec{n}}_{\ell,m_\ell}(k_\mathsf{cm}^2;L)$, defined as
\begin{align} \label{AppA:Eq:c}
c^{\vec{n}}_{\ell,m_\ell}(k_\mathsf{cm}^2;L)=\frac{\sqrt{4\pi}}{\gamma L^3}\bigg(\frac{2\pi}{L}\bigg)^{\!\ell-2}\,
Z^{\vec{n}}_{\ell,m_\ell}\bigg[1;\bigg(\frac{k_\mathsf{cm}L}{2\pi}\bigg)^2\,\bigg] \,, \quad
Z^{\vec{n}}_{\ell,m_\ell}[s;x^2]=\sum_{\vec{r}\in \mathcal{P}_{\vec{n}}} \frac{|\vec{r\,}|^\ell \, Y_{\ell m_\ell}(\vec{r})}{(|\vec{r\,}|^2-x^2)^s}\,,
\end{align}
where $\vec{n} = \frac{L}{2\pi} \vec{P}$, and $k_\cm$ is the $\cm$-frame momentum corresponding to $E_\cm$. Definitions of the boost-factor, $\gamma$, and the construction of the set of real vectors to be summed over, $\vec{r}\in \mathcal{P}_{\vec{n}}$, can be found in Ref.~\cite{Briceno:2014oea}. Using $Y^*_{\ell, m} = (-1)^{m} \, Y_{\ell, -m}$, it is easy to show that 
$c^{\vec{n}*}_{\ell,m_\ell} = c^{\vec{n}}_{\ell,-m_\ell}$. The presence of the factor $k_\cm^{-(\bar{\ell} +1)}$ in Eqn.~\ref{AppA:Eq:M} distinguishes the above and below threshold cases: above threshold $k_\cm$ is real and this factor is unchanged by complex conjugation, below threshold $k_\cm$ is imaginary, and complex conjugation can have an effect.

The integral over the product of three spherical harmonics can be expressed in terms of Clebsch-Gordan coefficients,
\begin{equation*}
\int\!\! d\Omega \; Y_{\ell m_\ell}^*  Y_{\bar{\ell} \bar{m}_\ell}^* Y^{}_{\ell' m_\ell'} = \sqrt{ \frac{(2\ell +1)(2\bar{\ell}+1)}{4\pi (2\ell'+1)} } 
\,\langle \ell m_\ell ; \bar{\ell} \bar{m}_\ell | \ell' m_\ell' \rangle \, \langle \ell 0; \bar{\ell} 0 | \ell'0\rangle,
\end{equation*}
and thus the integral is manifestly real. We leave it in the integral form as it more directly exposes the relevant symmetries.

Under complex conjugation of Eqn.~\ref{AppA:Eq:M} we can replace the resulting $c^{\vec{n}*}_{\ell,m_\ell}$ with $c^{\vec{n}}_{\ell,-m_\ell}$ and $Y_{\bar{\ell}, \bar{m}_\ell}$ with $(-1)^{\bar{m}_\ell} Y^*_{\bar{\ell}, -\bar{m}_\ell}$, and then noting that the sum over $\bar{m}_\ell$ is equivalent to a sum over $-\bar{m}_\ell$ obtain in the above threshold case where $k_\cm$ is real,
\begin{align}
\mathcal{M}^*_{\ell S J m a,\,  \ell' S' J' m' b} = \delta_{ab} \, \delta_{SS'}\sum_{\substack{m_\ell, m_\ell'\\ m_S}} \langle& \ell' m_\ell'; S m_S | J' m'\rangle\, \langle \ell m_\ell; S m_S | J m\rangle \, \nonumber \\
& \times \sum_{\bar{\ell}, \bar{m}_\ell}  \frac{(4\pi)^{3/2}}{k_\mathsf{cm}^{\bar{\ell} +1} } \, c^{\vec{n}}_{\bar{\ell}, \bar{m}_\ell}(k_\mathsf{cm}^2; L) \, \int\!\! d\Omega \; Y_{\ell' m'_\ell}^*  Y_{\bar{\ell} \bar{m}_\ell}^* Y^{}_{\ell m_\ell} \, ,
\end{align}
from which the symmetry under exchange of the totality of matrix indices is evident under a relabelling $m_\ell \leftrightarrow m_\ell'$, so that ${\mathcal{M}^*_{\ell S J m a,\,  \ell' S' J' m' b} = \mathcal{M}_{\ell' S' J' m' b,\, \ell S J m a} }$, and the hermiticity of $\bm{\mathcal{M}}$ above threshold is established. Subduction into irreducible representations of the relevant cubic symmetry groups does not affect this result as the subduction matrices are unitary.

\end{widetext}

\section{Avoided Eigenvalue Crossings}\label{AppB}

\begin{figure*}
	\centering
	\includegraphics[trim={0cm 0cm 0cm 0cm},clip,width=0.95\textwidth]{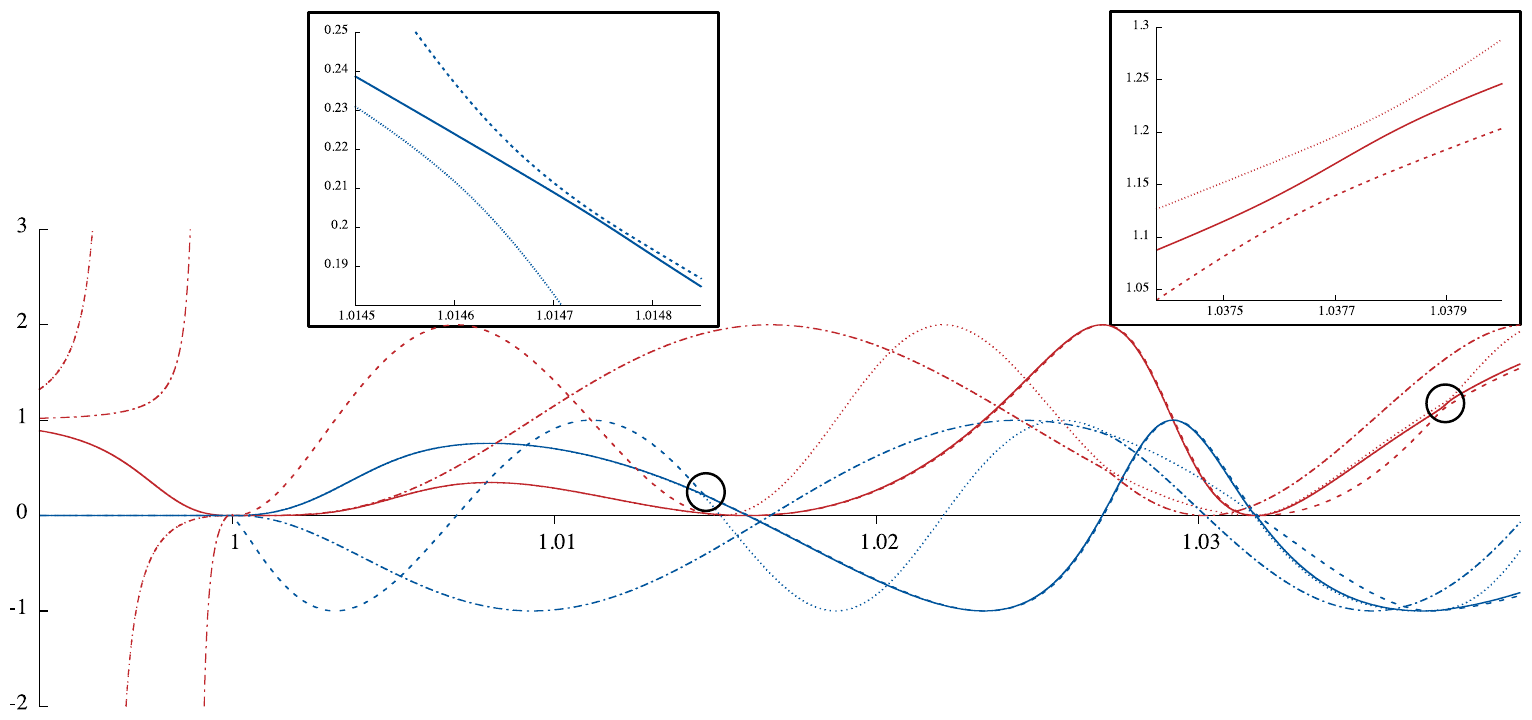}
	\caption{Eigenvalues, $\lambda_1,\dots,\lambda_4$ of $\bm{D}_{\bm{V}}(E_\cm)$. The real and imaginary components of each eigenvalue are shown in red and blue respectively and the different dashed lines correspond to the different eigenvalues. Highlighted are examples of avoided level crossings between the real and imaginary parts of different eigenvalues as discussed in the text.}
	\label{avoided}
\end{figure*}

Eigenvalues of $\bm{D_V}=\bm{1}+\bm{S}\cdot \bm{V}$ are observed to exhibit a behavior where they avoid crossing each other as $E_\cm$ is varied. Examples from the toy model considered in Section~\ref{SecIII} are shown in Figure~\ref{avoided}. This behaviour is in fact a general property that follows from the unitarity of the matrix $\bm{S}\cdot \bm{V}$ above threshold.

Defining unitary matrices at neighbouring energy points, ${\bm{W} = \bm{S}(E_\cm) \cdot \bm{V}(E_\cm)}$ and ${\bm{W}' = \bm{S}(E_\cm+ \Delta E_\cm) \cdot \bm{V}(E_\cm + \Delta E_\cm)}$, we can find {\em hermitian} matrices, $\bm{H}, \bm{H}'$ such that
\begin{align*}
\bm{W}&=\exp{i\bm{H}} \nonumber \\
\bm{W}'&=\exp{i\bm{H}'} ,
\end{align*}
and clearly the eigenvalues and eigenvectors of these matrices are trivially related,
\begin{align*}
\bm{H}\bm{v}^{(p)}&= \theta_p \, \bm{v}^{(p)} \nonumber \\
\bm{W}\bm{v}^{(p)}&= e^{i\theta_p}\, \bm{v}^{(p)}, 
\end{align*}
with $\theta_p$ being real, and simply related to the complex eigenvalues $\lambda_p$ of $\bm{D_V}$ as given in Eqn.~\ref{SecII:Eq:eigs}. The avoided level crossings that manifest in both the real and imaginary parts between some $\lambda_p$'s therefore appear also between the corresponding $\theta_p$'s.

By choosing to examine $\bm{H}$ we reduce the problem to one familiar in quantum mechanics: as $\bm{H}$ and $\bm{H}' - \bm{H}$ are hermitian this is the case considered in (degenerate) perturbation theory, where `avoided level crossings' are a generic feature whenever the perturbation ($\bm{H}' - \bm{H}$) connects the (degenerate) eigenstates. A classic illustrative example is the two-state case where in the \mbox{eigenbasis of $\bm{H}$},
\begin{equation*}
\bm{H}' = 
\begin{pmatrix} 
\theta_1 & 0 \\
0 & \theta_2
\end{pmatrix} +
\begin{pmatrix} 
P_{11} & P_{12} \\
P_{12}^* & P_{22} 
\end{pmatrix}\, ,
\end{equation*}
and the corresponding exact eigenvalues of $\bm{H}'$ are ${\theta'_1 = a + b}$, ${\theta'_2 = a - b}$, where
\begin{align*}
a &= \tfrac{1}{2} \big(\theta_1 + \theta_2 +P_{11} + P_{22} \big) \\
b &= \tfrac{1}{2}\sqrt{ \big(\theta_1 - \theta_2 + P_{11} - P_{22} \big)^2 + 4 \big|P_{12} \big|^2 } .
\end{align*}
It follows that unless both terms under the square root are zero, $\big|\theta'_2 - \theta'_1 \big| > 0$, and the eigenvalues do not cross.

\section{Tables of singularities}\label{AppC}

In this appendix, we record the singularities of $\bm{V}(E_\cm)$ in each irrep and volume relevant for the calculation presented in Sec.~\ref{SecIV}. We record singularities for irreps featuring subductions of $J^P=1^-$ and \emph{not} $J^P=1^+$ in Table~\ref{AppC:Tab:rho_irreps}, irreps featuring $J^P=1^+$ and \emph{not} $J^P=1^-$ in Table~\ref{AppC:Tab:b1_irreps}, and irreps featuring \emph{both} in Table~\ref{AppC:Tab:rho_b1_irreps}. For clarity, the singularities arising in different hadron-hadron channels (`$a$') is made explicit in each of the tables.

\begin{table}[h]
	\setlength{\tabcolsep}{2pt} 
	\small
	{\renewcommand{\arraystretch}{1.2}
		\begin{tabularx}{0.45\textwidth}{l @{\extracolsep{\fill}} llllll}
\hline
			$L/a_s$ & `$a$' & $[000]T_1^-$ & $[001]A_1$ & $[011]A_1$ & $[111]A_1$ & $[002]A_1$ \\
\hline
			%%%%%
			\multirow{3}{*}{$16$} &  $\pi\pi$ & 0.1250 & - &  - & -  & -  \\
			& $K\overline{K}$  & 0.1848 & - & -& -& - \\
			& $\pi\omega$ & 0.2151 & 0.2146 & 0.2175 & 0.2195 &  0.2146\\
			%%%%%
			\hline
\multirow{3}{*}{$20$} & $\pi\pi$ & 0.1299 & -  & - &  - & -\\
& $K\overline{K}$& 0.1881 & -& -& -& 0.1934 \\
& $\pi\omega$& 0.2186 & 0.2182 & 0.2201 & 0.2215 &  0.2182\\
			%%%%%
			\hline
\multirow{3}{*}{$24$} & $\pi\pi$ & 0.1324 &  - &  -&  - & 0.1381 \\
& $K\overline{K}$& 0.1900 & - &-&- & 0.1925 \\
& $\pi\omega$& 0.2204 & 0.2201 & 0.2215 & 0.2225 & 0.2202 \\
			\hline
			%%%%%
		\end{tabularx}
	}
	\caption{Singularities $E_\cm$ of the matrix $\bm{V}(E_\cm)$ as described in the text.}
	\label{AppC:Tab:rho_irreps}
\end{table}
%%%%%
\begin{table}[h]
	\setlength{\tabcolsep}{2pt} 
	\small
	{\renewcommand{\arraystretch}{1.2}
		\begin{tabularx}{0.45\textwidth}{l @{\extracolsep{\fill}} llllll}
			\hline
			$L/a_s$ & `$a$' & $[000]T_1^+$ & $[001]A_2$ & $[011]A_2$ & $[111]A_2$ & $[002]A_2$ \\
			\hline
			%%%%%
			\multirow{2}{*}{$16$} & $\pi\omega$ &0.2151$\{2\}$& 0.2146 & 0.2175 & 0.2195 & 0.2146 \\
			& $\pi\phi$ & 0.2396 & 0.2459 & - & - & 0.2481 \\
			%%%%%
			\hline
			\multirow{2}{*}{$20$} & $\pi\omega$ &0.2186$\{2\}$& 0.2182 & 0.2201 & 0.2215 & 0.2182 \\
			& $\pi\phi$& 0.2429 & 0.2471 & - & - & 0.2485 \\
			%%%%%
			\hline
			\multirow{2}{*}{$24$} & $\pi\omega$ & 0.2204$\{2\}$& 0.2201 & 0.2215 & 0.2225 & 0.2202 \\
			& $\pi\phi$& 0.2446 & 0.2476 & - & - & - \\
			\hline
			%%%%%
		\end{tabularx}
	}
	\caption{As in Table~\ref{AppC:Tab:rho_irreps}. Integers in curly parenthesis denote multiplicities of the singularities if greater than one.}
	\label{AppC:Tab:b1_irreps}
\end{table}
%%%%%
\begin{table}[h]
	\setlength{\tabcolsep}{2pt} 
	\small
	{\renewcommand{\arraystretch}{1.2}
		\begin{tabularx}{0.4\textwidth}{l @{\extracolsep{\fill}} lllll}
			\hline
			$L/a_s$ & `$a$' & $[001]E_2$ & $[011]B_1$ & $[011]B_2$ & $[111]E_2$ \\
			\hline
			%%%%%
			\multirow{4}{*}{$16$} & $\pi\pi$ &0.1242 & 0.1363 & 0.1236 & 0.1311 \\
			& $K\overline{K}$ & 0.1842& - & 0.1838  & 0.1900 \\
			& $\pi\omega$ & 0.2146  & 0.2141 & 0.2229 & 0.2195 \\
			& $\pi\phi$ & 0.2459 & -& -& - \\
			%%%%%
			\hline
			\multirow{4}{*}{$20$} & $\pi\pi$ &0.1294 & -& 0.1290 & 0.1342 \\
			& $K\overline{K}$ & 0.1878 & -& 0.1874 & 0.1919 \\
			& $\pi\omega$ & 0.2182 & 0.2179 & 0.2238 & 0.2215 \\
			& $\pi\phi$ & 0.2471 &- & -& - \\
			%%%%%
			\hline
			\multirow{4}{*}{$24$} & $\pi\pi$ &0.1321& - & 0.1318 & 0.1358 \\
			& $K\overline{K}$ & 0.1896 &- & 0.1894 & 0.1928 \\
			& $\pi\omega$ & 0.2201 & 0.2199 & 0.2242 & 0.2225 \\
			& $\pi\phi$ & 0.2476 & -& -& -\\
			\hline
			%%%%%
		\end{tabularx}
	}
	\caption{As in Table~\ref{AppC:Tab:rho_irreps}.}
	\label{AppC:Tab:rho_b1_irreps}
\end{table}

%%%%% BIBLIO %%%%%

\bibliographystyle{apsrev4-1}
\bibliography{bib.bib}

\end{document}